\documentclass[11pt,a4paper]{article}
\usepackage{amsmath,latexsym,amssymb,amsfonts}
\usepackage[dvips]{graphicx}
\usepackage{bm}

\begin{document}
\title{\textbf{Black hole complementarity and firewall in\\ two dimensions}}
\author{Wontae Kim$^{a,b}$\footnote{wtkim@sogang.ac.kr},\; Bum-Hoon Lee$^{a,b}$\footnote{bhl@sogang.ac.kr}\; and Dong-han Yeom$^{a,c}$\footnote{innocent.yeom@gmail.com}\\
\textit{$^{a}$\small{Center for Quantum Spacetime, Sogang University, Seoul 121-742, Republic of Korea}}\\
\textit{$^{b}$\small{Department of Physics, Sogang University, Seoul 121-742, Republic of Korea}}\\
\textit{$^{c}$\small{Yukawa Institute for Theoretical Physics, Kyoto University, Kyoto 606-8502, Japan}}
}
\maketitle

\begin{flushright}
{\tt  YITP-13-33}
\end{flushright}

\begin{abstract}
In connection with black hole complementarity, we study the possibility of the duplication of information in the RST model which is an exactly soluble quantized model in two dimensions. We find that the duplication of information can be observed without resort to assuming an excessively large number of scalar fields. If we introduce a firewall, then we can circumvent this problem; however, the firewall should be outside the event horizon.
\end{abstract}

\newpage

\tableofcontents

\newpage

\section{Introduction}

The information loss problem in evaporating black holes \cite{Hawking:1976ra} has been an important problem that should be resolved by the theory of quantum gravity, so that there have been many approaches
in order to understand and resolve this problem along with string theory.
In particular, the efforts to resolve the information loss problem appear in the form of
so called \textit{black hole complementarity} \cite{Susskind:1993if}. The principle of black hole complementarity is closely related to the membrane paradigm \cite{Thorne:1986iy}, the D-brane picture \cite{Callan:1996dv}, and the AdS/CFT correspondence \cite{Lowe:1999pk}.

According to black hole complementarity, two observers around a black hole are complementary: an in-going observer who observes classical metric and eventually touches a physical singularity, while an asymptotic observer observes semi-classical Hawking radiation and eventually gather all the information from Hawking radiation. The question is whether these two observers are consistent or not. Perhaps, if two observers can meet each other, then there will be a contradiction of this complementary picture.
This issue has been discussed by Susskind and Thorlacius in detail \cite{Susskind:1993mu}. Such two copies of information cannot be observed by a single observer and hence complementarity seems to be consistent. The main point is that the outside observer should wait until the information retention time to see a piece of information \cite{Page:1993df,Page:1993wv}. This makes the observation of the duplication of information improbable. Furthermore, the complementarity principle looks marginally safe even
though we consider a shorter time scale, so-called the scrambling time \cite{Hayden:2007cs}.

However, there have been some authors to against these safety arguments \cite{Yeom:2008qw,Hong:2008mw,Hong:2008ga}. The main point was that if we include a lot of scalar fields that contribute to Hawking radiation, then the duplication of information can be observed. Although other authors suggested a similar possibility \cite{Dvali:2007hz}, the black hole could not be semi-classical. On the other hand, in Ref.~\cite{Yeom:2009zp}, a large $N$ rescaling with a semi-classical black hole was considered. According to this work, regarding the information retention time, we need the
exponentially large number of scalar fields to violate the complementarity principle, while we need a reasonably small number of scalar fields regarding the scrambling time. Moreover, recently, Almheiri, Marolf, Polchinski and Sully (AMPS) \cite{Almheiri:2012rt} could find a problem in the original version of complementarity from the other point of view. They introduced so-called the firewall to overcome the problem. Now there is an interesting controversy in this field \cite{Susskind:2012rm,Mathur:2012jk,Bousso:2012as,Hwang:2012nn}.

Regarding the information retention time, the required number of scalar fields was too large to accept easily although there is no fundamental limitation to assume such a finite number of scalar fields. This could be overestimated result due to a crude approximation, since it is quite difficult to precisely measure the time difference inside the evaporating black hole. On the other hand, in two dimensional cases of the Callan-Giddings-Harvey-Strominger (CGHS) model and the Russo-Susskind-Thorlacius (RST) model
\cite{Callan:1992rs,deAlwis:1992hv,Bilal:1992kv,Russo:1992ax}, we know more on an evaporating black hole that is motivated from the study of the back reaction of the geometry.
Subsequently, there have been many gravitational applications in terms of
these models along with some modifications \cite{Kim:1995wr}.
If we can find a scaling behavior in two dimensions, it will make our calculations and the possibility of the duplication experiment clearer. We may be able to calculate the exact number of scalar fields that is required to violate the principle of complementarity. Eventually, we can judge whether the required number of scalar fields can be acceptable or not. In this paper, we would like to study the possibility of the duplication experiment mainly in the context of the RST model.
In Sec.~\ref{sec:pre}, we first define the technical details of the duplication experiment and summarize previous results on the large $N$ rescaling. In Sec.~\ref{sec:two}, we recapitulate the two dimensional black hole models and calculate some important quantities to study the duplication experiment. In Sec.~\ref{sec:dup}, we show that the duplication experiment is indeed possible by defining a scaling law for two dimensional version. Moreover, we will find the criterion of the required number of scalar fields that violates the principle of black hole complementarity. Finally, in Sec.~\ref{sec:con}, we summarize and comment on other possible alternatives, including the firewall proposal and the fuzzball conjecture.

\section{\label{sec:pre}Preliminaries}
In this section, we review the principle of black hole complementarity \cite{Susskind:1993if} and the duplication experiment \cite{Susskind:1993mu}. In addition, we discuss previous arguments that violate black hole complementarity, so-called the large $N$ rescaling \cite{Yeom:2009zp}.
This will make the purpose of this paper clear.

\subsection{Black hole complementarity}

First of all, it is worthwhile to illustrate the original idea of black hole complementarity. Black hole complementarity assumes the five contents:
\begin{description}
\item[Assumption 1] Unitarity, and hence the loss or copy of information should not be observed by any local observer.
\item[Assumption 2] For an asymptotic observer, the unitary local quantum field theory is a good description.
\item[Assumption 3] For an in-going observer, general relativity is a good description.
\item[Assumption 4] The area of the black hole is proportional to the coarse-grained entropy of the black hole. Hence, information comes out around the information retention time $\sim M^{3}$ \cite{Page:1993df,Page:1993wv}.
\item[Assumption 5] There is an ideal observer who can control and count information from Hawking radiation.
\end{description}
The motivation of black hole complementarity is to make these assumptions consistent.

\begin{figure}
\begin{center}
\includegraphics[scale=0.7]{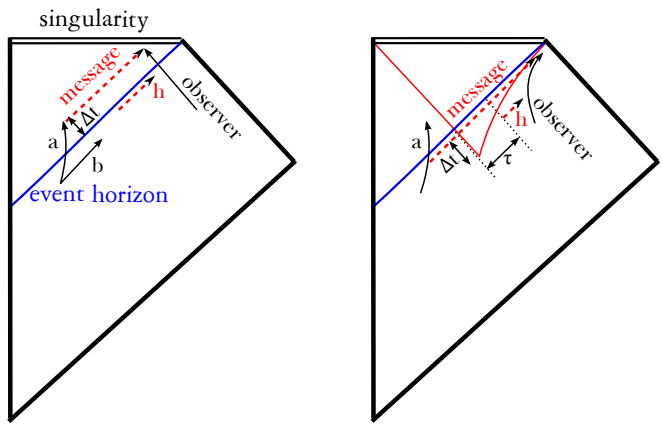}
\caption{\label{fig:Schwarzschild_duplication}Left: Conceptual description of duplication experiment. Right: Our stronger version of duplication experiment to see the duplication of information outside the event horizon. Here, $h$ means the information about $a$ that is copied by Hawking radiation.}
\end{center}
\end{figure}

At the first glimpse, this is not consistent, since there should be a copy of information for an asymptotic observer and an in-going observer. However, if no one can see both copies of information, then there will be no problem. We first define the duplication experiment and then
 discuss why it seemed to be impossible by previous authors \cite{Susskind:1993mu}. The logical sketch of the duplication experiment is given by the following steps, which is shown in Fig.~\ref{fig:Schwarzschild_duplication}.
\begin{itemize}
\item[1.] Create an entangled spin pair $a$ and $b$.
\item[2-1.] An in-going observer $\mathcal{A}$ brings $a$ into the black hole, while $b$ is outside.
\item[2-2.] $\mathcal{A}$ sends a signal on $a$ along the out-going direction.
\item[3-1.] An observer $\mathcal{O}$ who is outside the black hole measures $b$ and he/she knows the state of $b$.
\item[3-2.] After the information retention time, Hawking radiation emits the quantum information on $a$: we call this $h$.
\item[3-3.] The observer $\mathcal{O}$ measures $h$ and he/she knows that it is on $a$ by comparing $b$.
\item[4.] The observer $\mathcal{O}$ falls into the black hole and fortunately sees the signal of $\mathcal{A}$ on $a$.
\item[5.] The observer $\mathcal{O}$ knows that $a$ is the original information by comparing $b$.
\end{itemize}
If it is possible, then the observer $\mathcal{O}$ sees a duplication of information. Therefore, the observer $\mathcal{O}$ experiences a kind of non-unitary evolution, so that it is inconsistent with the above assumptions.

Where is the loophole?
According to Susskind and Thorlacius \cite{Susskind:1993mu}, Step 2-2 and Step 4 are inconsistent. The observer $\mathcal{O}$ should fall into the black hole \textit{after} the information retention time $\tau \sim M^{3}$. Then the observer $\mathcal{A}$ should send a signal between the time
\begin{eqnarray}
\Delta t \simeq \exp \left( - \frac{\tau}{r_{\mathrm{h}}} \right)
\end{eqnarray}
where $r_{\mathrm{h}}\sim M$ is the horizon. In addition, to attach a bit of information, one has to satisfy the uncertainty principle as
\begin{eqnarray}
\Delta t \Delta E \sim \hbar = 1,
\end{eqnarray}
which yields $\Delta  E \simeq \exp M^2$.
As long as $\Delta E > M$, then such an event cannot happen.
We call the following inequality the \textit{consistency condition}:
\begin{eqnarray}
\Delta E > M.
\end{eqnarray}
However, if the consistency condition does not hold, then this shows the inconsistency of the assumptions of black hole complementarity.

\subsection{Large $N$ rescaling in dimensions $D>2$}
There have been some arguments against this resolution of black hole complementarity.
We first summarize the results of large $N$ rescaling \cite{Yeom:2009zp} and show that it allows the duplication experiment. Let us consider the case such that
we can build a classical black hole with a single scalar field, while it evaporates by $N$ scalar fields. Then this satisfies the semi-classical equations
\begin{eqnarray}
G_{\mu\nu} = 8\pi \left(T_{\mu\nu} + \hbar N \langle T_{\mu\nu}\rangle\right)
\end{eqnarray}
up to the one loop order. By fixing $G=c=N\hbar = 1$,
we obtain a solution of the equations.
We already fixed $c$ and $G$ and hence all dimensional quantities are determined by the unit length $l_{\mathrm{Pl}}$. However, since the Plank constant of $\hbar = 1/N = l_{\mathrm{Pl}}^{D-2}$ is not fixed, the unit length is free. As $N$ increases, $\hbar$ decreases, and hence in general, the unit length $l_{\mathrm{Pl}}$ decreases by factor $N^{-1/(D-2)}$ \cite{Hong:2008mw}. Therefore, the physical size increases in Planck units.
For a given quantity $X$ with a length dimension $[X]=L^{\alpha}$ that is a solution with $N=1$, if we transform \textit{all} possible classical quantities by the following scaling law
\begin{eqnarray}
X' = X N^{\alpha/(D-2)},
\end{eqnarray}
then $X'$ is also a solution of large $N$ semi-classical equations \cite{Yeom:2009zp}.
Note that it is not well-defined for two dimensional case, which will be discussed in later.

If the large $N$ rescaling is allowed, then the duplication experiment becomes possible.
The time difference and the mass should be rescaled as follows:
\begin{eqnarray}
\Delta t \rightarrow N^{1/(D-2)} \Delta t, \;\;\;\; M \rightarrow N^{(D-3)/(D-2)} M,
\end{eqnarray}
and from the uncertainty relation,
\begin{eqnarray}
\Delta E = N^{-1/(D-2)} \frac{1}{\Delta t}.
\end{eqnarray}
The consistency relation for black hole complementarity satisfying $\Delta E > M$ is
\begin{eqnarray}
N^{-1/(D-2)} \frac{1}{\Delta t} > N^{(D-3)/(D-2)} M.
\end{eqnarray}
Therefore, if
\begin{eqnarray}
N > \frac{1}{M\Delta t},
\end{eqnarray}
then $\Delta E <M$, and hence the observation of the duplication is allowed, where $M$ and $\Delta t$ are the mass and the time difference that measured in $N=1$ case.

\subsection{Motivations}

The above argument reveals the potential inconsistency of black hole complementarity. However, there are some questions. First, is there any exact calculation for $\Delta t$ and $N$ to violate black hole complementarity, while Susskind and Thorlacius \cite{Susskind:1993mu} calculated approximated way? Second, why is there no such scaling behavior in two dimensions and can there be any analogous in two dimensions? Third, what is the minimum required number of scalar fields to violate black hole complementarity?

These questions will be resolved by observing two dimensional evaporating black hole models \cite{Callan:1992rs,deAlwis:1992hv,Bilal:1992kv,Russo:1992ax}. In this paper, we restrict Step 2-2 such that $\mathcal{A}$ sends a signal inside the apparent horizon and outside the event horizon (Right of Fig.~\ref{fig:Schwarzschild_duplication}). Then we can modify Step 4 such that the observer $\mathcal{O}$ sees the signal $a$ outside the event horizon (see also \cite{Hwang:2012nn}). This is a stronger experiment than the previous one and hence if the duplication experiment in this setup is successful even with this restriction, then there is no hope to prohibit the duplication experiment. We will further show that such event is possible with a reasonable number of scalar fields. All calculations are definite since we deal with an exactly solvable model in two dimensions.

\section{\label{sec:two}Two dimensional black hole models}

\subsection{Models}
We begin with the two dimensional dilatonic black hole model \cite{Callan:1992rs,deAlwis:1992hv,Bilal:1992kv,Russo:1992ax}.
In the conformal gauge, the metric as a function of the advanced time $x^{+}$ and the retarded time $x^{-}$
is written as
\begin{eqnarray}
ds^{2} = - e^{2\rho} dx^{+}dx^{-}.
\end{eqnarray}
The following two dimensional dilatonic black hole model has been suggested by Callan, Giddings, Harvey and Strominger \cite{Callan:1992rs},
\begin{eqnarray}
S_{\mathrm{CGHS}} = \frac{1}{\pi} \int d^{2} x \left[ e^{-2\phi} \left( -2\partial_{+}\partial_{-}\rho + 4 \partial_{+}\phi\partial_{-}\phi - \lambda^{2} e^{2\rho} \right) - \frac{1}{2} \sum_{i=1}^{N} \partial_{+}f_{i}\partial_{-}f_{i} + \kappa \hbar \partial_{+}\rho\partial_{-}\rho \right],
\end{eqnarray}
where $\phi,~\lambda,~f_{i}$  are the dilaton field, the cosmological constant, scalar fields, and $\kappa \equiv N/12$ is proportional to the number of scalar fields.
Note that it is not easy to solve this model by analytic functions and hence we need to use numerical techniques \cite{PS}.
On the other hand, if we impose further symmetry, then we can find an exactly solvable model. For example, de Alwis \cite{deAlwis:1992hv} and Billal and Callan \cite{Bilal:1992kv} introduced a further symmetry, although there seems to be a problem. More interesting exactly solvable model has been suggested by Russo, Susskind and Thorlacius \cite{Russo:1992ax}:
\begin{eqnarray}
S_{\mathrm{RST}} = S_{\mathrm{CGHS}} + \frac{\kappa \hbar}{\pi} \int d^{2} x \; \phi \partial_{+}\partial_{-}\rho.
\end{eqnarray}
The extra term comes from an ambiguity of the trace anomaly, which yields exact solubility.
We will mainly study the RST model, and compare the final results with the CGHS model.

\subsection{RST exact solution}
The exact solution can be obtained by defining two functions \cite{Russo:1992ax},
\begin{eqnarray}
\Omega = \frac{\sqrt{\kappa}}{2}\phi + \frac{e^{-2\phi}}{\sqrt{\kappa}}, \;\;\;\; \chi = \sqrt{\kappa}\rho - \frac{\sqrt{\kappa}}{2}\phi + \frac{e^{-2\phi}}{\sqrt{\kappa}},
\end{eqnarray}
which is written as
\begin{eqnarray}
\Omega = \chi = -\frac{\lambda^{2}x^{+}x^{-}}{\sqrt{\kappa}} - \frac{\sqrt{\kappa}}{4} \log \left( -\lambda^{2} x^{+}x^{-}\right) - \frac{M}{\lambda \sqrt{\kappa} x_{0}^{+}} \left(x^{+}-x_{0}^{+}\right) \theta \left(x^{+}-x_{0}^{+}\right),
\end{eqnarray}
where a single shock wave at $x^{+}=x_{0}^{+}$ is introduced in order to form a black hole.

\begin{figure}
\begin{center}
\includegraphics[scale=0.4]{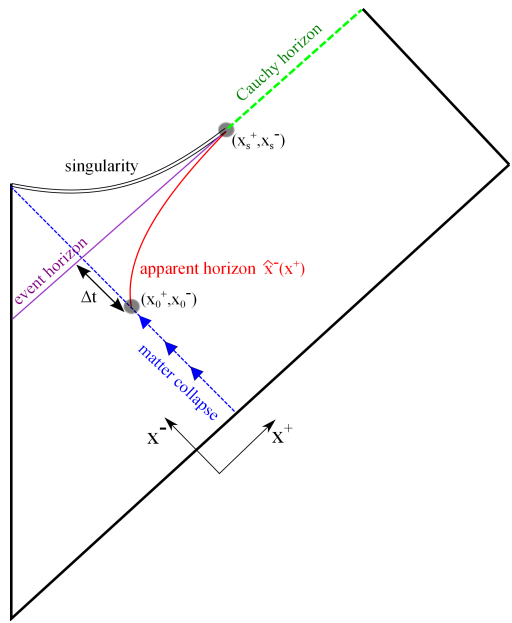}
\caption{\label{fig:structure}Causal structure of RST model.}
\end{center}
\end{figure}

The location of the apparent horizon $(\hat{x}^{+}, \hat{x}^{-})$ can be written by the relation,
\begin{eqnarray}
\hat{x}^{+} = -\frac{\kappa}{4\lambda^{2}} \frac{1}{\hat{x}^{-}+M/\lambda^{3}x_{0}^{+}}.
\end{eqnarray}
Therefore, when the black hole is formed around ${x}_{0}^{+}$, the apparent horizon becomes
\begin{eqnarray}
x_{0}^{-} = - \left( \frac{M}{\lambda^{3}x_{0}^{+}} + \frac{\kappa}{4 \lambda^{2} x_{0}^{+}} \right).
\end{eqnarray}
Next, we can calculate the endpoint of the evaporation $(x_{\mathrm{s}}^{+}, x_{\mathrm{s}}^{-})$ as
\begin{eqnarray}
x_{\mathrm{s}}^{+} &=& \frac{\kappa \lambda x_{0}^{+}}{4M} \left( e^{4M/\kappa\lambda} - 1 \right),\\
x_{\mathrm{s}}^{-} &=& -\frac{M}{\lambda^{3}x_{0}^{+}} \frac{1}{1 - e^{-4M/\kappa\lambda}}.
\end{eqnarray}
As a result, the causal structure is given in Fig.~\ref{fig:structure}.
The description is reliable as long as $N \gg 1$ even in this one-loop approximation.
To find the corresponding semi-classical black hole in $4$-dimensional limit, we should further assume $Z \equiv M/\kappa \gg 1$.

\subsection{Time difference between the apparent horizon and the event horizon} Along the initial in-going surface $x^{+}=x_{0}^{+}$, the difference of the areal radius between the apparent horizon and the event horizon $\Delta R \equiv R_{\mathrm{apparent\; horizon}} - R_{\mathrm{event\; horizon}}$ becomes
\begin{eqnarray}
\Delta R = \left.\left( e^{-2\phi} + \frac{\kappa}{2}\phi \right)\right|_{\mathrm{apparent\; horizon}} - \left.\left( e^{-2\phi} + \frac{\kappa}{2}\phi \right)\right|_{\mathrm{event\; horizon}}.
\end{eqnarray}
Slightly before the thin shell, the space-time corresponds the flat space, and at that time, the difference of the areal radius corresponds the time that the light moves between two points. We think that the event to send a signal to out-going direction can happen at the very early stage of the gravitational collapse.
Then, we can reasonably guess that the time difference between two points is $\Delta t = \Delta R$.
In this model, we can exactly calculate $\Delta R$ or $\Delta t$, which is
\begin{eqnarray}
\label{time}
\Delta t = \lambda^{2} x_{0}^{+} \left| x_{0}^{-} - x_{s}^{-} \right| - \frac{\kappa}{4} \log \frac{x_{0}^{-}}{x_{s}^{-}}.
\end{eqnarray}
Note that
\begin{eqnarray}
\lambda^{2} x_{0}^{+} \left| x_{0}^{-} - x_{\mathrm{s}}^{-} \right| = \frac{\kappa}{4} + \frac{M}{\lambda}\left[ 1 - \left( 1 - e^{-4M/\kappa\lambda} \right)^{-1} \right]
\simeq \frac{\kappa}{4}
\end{eqnarray}
and
\begin{eqnarray}
\frac{\kappa}{4} \log \frac{x_{0}^{-}}{x_{\mathrm{s}}^{-}} = \frac{\kappa}{4} \left[ \log \left(1+\frac{\kappa\lambda}{4M} \right) + \log \left(1- e^{-4M/\kappa\lambda} \right) \right]
\ll \frac{\kappa}{4}
\end{eqnarray}
if $M/\kappa \gg 1$.
Therefore,
\begin{eqnarray}
\Delta t &=& \frac{\kappa}{4} + \frac{M}{\lambda}\left[ 1 - \left( 1 - e^{-4M/\kappa\lambda} \right)^{-1} \right] - \frac{\kappa}{4} \left[ \log \left(1+\frac{\kappa\lambda}{4M} \right) + \log \left(1- e^{-4M/\kappa\lambda} \right) \right]\\
&\simeq& \frac{\kappa}{4} \label{anothertime}
\end{eqnarray}
for the semi-classical black hole with the condition $M/\kappa \gg 1$.
Note that this result is quite universal in two dimensional models in the sense that for the CGHS model \cite{Susskind:1992gd} the location of the horizon $\hat{x}^{-}(x^{+})$ is approximately
\begin{eqnarray}
\hat{x}^{-}(x^{+}) \simeq \hat{x}^{-}(x_{0}^{+}) + \frac{\kappa}{4\lambda^{2}} \left( \frac{1}{x_{0}^{+}} - \frac{1}{x^{+}} \right).
\end{eqnarray}
Then, the proper time difference between the event horizon and the initial apparent horizon is
\begin{eqnarray}
\Delta t \simeq \lambda^{2} x_{0}^{+} \left| \hat{x}^{-}(x^{+}) - \hat{x}^{-}(x_{0}^{+}) \right| \simeq \frac{\kappa}{4} \left.\left( 1 - \frac{x_{0}^{+}}{x^{+}} \right)\right|_{x^{+}\gg x_{0}} \simeq \frac{\kappa}{4}.
\end{eqnarray}
Here, we choose $x^{+}\gg x_{0}$, since the event horizon is determined for the limit of $x^{+}\gg x_{0}$.
In this respect, $\Delta t \simeq \kappa/4$ is quite general and correct result in two dimensional evaporating black hole models.

\section{\label{sec:dup}Duplication experiment}

\subsection{Scaling behaviors in two dimensions}
The RST action as well as CGHS model is invariant under the following transformation,
\begin{eqnarray}
\kappa &\rightarrow& \alpha \kappa,\\
\rho &\rightarrow& \rho - \frac{1}{2} \log \alpha,\\
\phi &\rightarrow& \phi - \frac{1}{2} \log \alpha,\\
x^{-} &\rightarrow& \alpha x^{-},\\
M &\rightarrow& \alpha M,
\end{eqnarray}
where $\alpha>1$ for convenience.
Note that the similar scaling transformation was introduced in Ref.~\cite{Ori:2010nn},
but this is more advanced version.
The invariance can be easily checked with the help of
\begin{eqnarray}
\Omega \rightarrow \sqrt{\alpha} \Omega - \frac{\sqrt{\alpha\kappa}}{4} \log \alpha, \;\;\;\; \chi \rightarrow \sqrt{\alpha} \chi - \frac{\sqrt{\alpha\kappa}}{4} \log \alpha.
\end{eqnarray}
The rescaled values satisfy the equations of motion. The important results are
\begin{eqnarray}
\Delta t \rightarrow \alpha \Delta t, \;\;\;\; M \rightarrow \alpha M.
\end{eqnarray}

\subsection{Application to the duplication experiment}
For the case of $\alpha=1$,
the consistency of black hole complementarity implies that
\begin{eqnarray}
\Delta E_{\alpha=1} \simeq \frac{\hbar}{\Delta t_{\alpha=1}} > M_{\alpha=1},
\end{eqnarray}
where $\Delta E_{\alpha=1}$ and $ M_{\alpha=1}$ are fixed values.
After rescaling, if black hole complementarity still holds, then
\begin{eqnarray}
\Delta E_{\alpha>1} \simeq \frac{\hbar}{\Delta t_{\alpha>1}} > M_{\alpha>1}
\end{eqnarray}
with the rescaling transformations of
\begin{eqnarray}
\Delta t_{\alpha>1} &=& \alpha \Delta t_{\alpha=1},\\
M_{\alpha>1} &=& \alpha M_{\alpha=1},\\
\Delta E_{\alpha>1} &  \simeq & \frac{\hbar}{\Delta t_{\alpha>1}} = \frac{\hbar}{\alpha\Delta t_{\alpha=1}}.
\end{eqnarray}
It means that the consistency condition should be
\begin{eqnarray}
\Delta E_{\alpha=1} > \alpha^{2} M_{\alpha=1}.
\end{eqnarray}
For given $\Delta E_{\alpha=1}$ and $M_{\alpha=1}$ in the evaporating black hole solution,
 there exists a special $\alpha$ such that the consistency condition of black hole complementarity is violated, i.e., $\Delta E_{\alpha=1} < \alpha^{2} M_{\alpha=1}$.
Note that this rescaling makes all kinds of duplication experiment (whatever outside or inside the event horizon, whatever the information retention time or scrambling time) possible as long as we have sufficiently large $\alpha$.

One further note for careful calculations is that if we do the duplication experiment, then the experiment in itself can perturb the RST solution. Even though the solution allows the duplication experiment, the realistic application needs careful considerations. One can set a bound such that if the signal requires the energy less than $\epsilon M \ll M$, then one can still trust the causal structure of the exact solution without any further back reaction of the geometry.
Finally, the condition for a successful duplication experiment is given by
\begin{eqnarray}
\label{breaking}
\Delta E_{\alpha=1} < \alpha^{2} \epsilon M_{\alpha=1}.
\end{eqnarray}

\begin{figure}
\begin{center}
\includegraphics[scale=0.4]{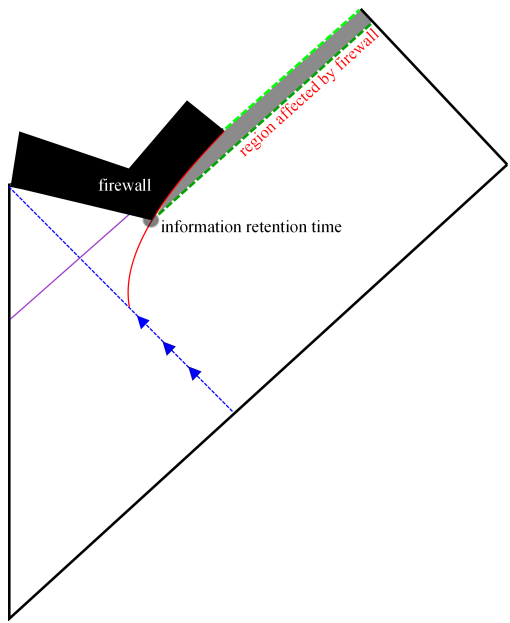}
\caption{\label{fig:firewall}The location of the firewall to prohibit the duplication of information. To prevent the duplication experiment, the firewall should be outside the event horizon. Hence, there is a region outside the black hole where is affected by the firewall.}
\end{center}
\end{figure}

\subsection{How large $\kappa$?}
We consider a black hole with a given $Z \equiv M/\kappa$ which
is invariant up to the rescaling. We already calculated $\Delta t$ in Eqs.~\eqref{time} and \eqref{anothertime}. The condition Eq.~\eqref{breaking} to violate black hole complementarity is written in the form of
\begin{eqnarray}
\alpha > \frac{1}{\sqrt{\Delta t_{\alpha=1}M_{\alpha=1}}}.
\end{eqnarray}
So the required $\kappa_{\alpha>1}$ is
\begin{eqnarray}
\kappa_{\alpha>1} = \alpha \kappa_{\alpha=1} > \frac{\kappa_{\alpha=1}}{\sqrt{\Delta t_{\alpha=1}M_{\alpha=1}}} = \sqrt{Z\frac{\kappa_{\alpha=1}}{\Delta t_{\alpha=1}}}.
\end{eqnarray}
In our model, $\kappa_{\alpha=1}/\Delta t_{\alpha=1} \simeq 4$
and hence the required scalar fields are
\begin{eqnarray}
\kappa_{\alpha>1} \geq 2 Z^{1/2}.
\end{eqnarray}
If we further consider the $\epsilon$ effect, then
\begin{eqnarray}
\kappa_{\alpha>1} \geq 2 \left(\frac{Z}{\epsilon}\right)^{1/2}.
\end{eqnarray}
The required number of scalar fields can be reasonably small.
For example, for a given black hole $M \sim 10^{6}$ with the limit of consistency $\epsilon \sim 10^{-6}$ (hence, $\epsilon M \sim 1$), if $\kappa \sim 10^{2}$ (hence, $M/\kappa \sim 10^{4} \gg 1$), then $\Delta t \sim 10^{2}$ and $\Delta E \sim 10^{-2}$. Therefore, this easily violates the consistency condition $\Delta E > \epsilon M$.

\subsection{Where is the firewall?} AMPS introduced the firewall around the event horizon \cite{Almheiri:2012rt,Susskind:2012rm} to maintain the basic philosophy of complementarity,
which amounts to dropping Assumption 3. However, as we discussed in this paper, the duplication experiment is possible even outside the event horizon. Therefore, the firewall should be, if it exists, located around the apparent horizon which is outside the event horizon after the information retention time, as shown in Fig.~\ref{fig:firewall} \cite{Hwang:2012nn}. This is the only way to maintain the basic principles of black hole complementarity. Then, this can imply that we should also drop Assumption 2 at the same time, since the firewall outside the event horizon should affect to an asymptotic observer (gray colored region in Fig.~\ref{fig:firewall}), where such an effect seems not semi-classical. If one does not like this possibility, then we should modify other assumptions, e.g., Assumptions 4 or 5.

\section{\label{sec:con}Conclusion}
We have investigated the consistency of black hole complementarity in two dimensions
in terms of dilatonic black holes. We could use the exact solutions as well as approximate results to calculate the precise condition to do the duplication experiment. The conclusion is worthwhile to highlight in that the duplication is possible using a \textit{reasonable} number (not exponentially large) of scalar fields that can be allowed.

Our conclusion is quite strong to restrict previous opinions on the information loss problem. For example, to keep the original philosophy of black hole complementarity, AMPS introduced the firewall \cite{Almheiri:2012rt}; however, we can further restrict that the firewall should be outside the event horizon and hence there should be an effect from the firewall to the asymptotic observer. Also, our results can give further requests to the fuzzball picture \cite{Mathur:2012jk}. If the fuzzball picture allows free-infall into the black hole, then the observation of the duplication of information should be allowed. To make the fuzzball picture more consistently, the picture should not allow free-infall or should give further stronger restriction to the in-falling observer. The duplication experiment is even possible outside the event horizon, and hence if a bit of information is duplicated at once, even though it takes a long time to read the information \cite{HH}, the duplication will be eventually observed by an asymptotic observer.

The most important question is this: \textit{which assumption is wrong among the five listed assumptions?} In this paper, we cannot answer to this question. If we drop Assumption 3, then we should drop Assumption 2 at the same time. We think that there is no evidence to do this way. Perhaps, to drop Assumption 4 or Assumption 5 can be a viable alternative, although we should be careful.

\newpage

\section*{Acknowledgment}
WK would like to thank M. S. Eune, Y. Kim,  E. J. Son for exciting discussions and DY would like to thank D. Hwang for helpful discussions. DY, WK and BHL are supported by the National Research Foundation of Korea(NRF) funded by the Korea government(MEST, 2005-0049409) through the Center for Quantum Spacetime(CQUeST) of Sogang University. WTK was supported by the Sogang University Research Grant of 2013(201310022). DY was supported by the JSPS Grant-in-Aid for Scientific Research (A) No. 21244033.

\end{document}